\documentclass[sigplan, 10pt]{acmart}

\renewcommand\footnotetextcopyrightpermission[1]{}
\AtBeginDocument{%
  }
\setcopyright{none}
\usepackage{subcaption}
\usepackage{algorithm}
\usepackage{algpseudocode}
\usepackage{graphicx}
\usepackage{subcaption}
\usepackage{booktabs}
\usepackage{multirow}
\usepackage{xcolor}
\usepackage{enumitem}
\usepackage{pifont}
\usepackage{tcolorbox}
\usepackage{hyperref}
\usepackage{url}
\usepackage{lipsum}

\usepackage{tikz}

\newcommand{\blackcircled}[1]{%
\tikz[baseline=(char.base)]{
  \node[shape=circle,fill=black,inner sep=1pt,minimum size=10pt] (char) {\color{white}\footnotesize #1};
}}
\begin{document}
\pagestyle{plain}
\title[SwiftCache: LLM Serving for Multi-turn Conversations with Heterogeneous KV Cache Sharing]{SwiftCache: Efficient LLM Serving for Multi-turn Conversations with Heterogeneous KV Cache Sharing}

% \author{Jianmin Hu}
% \email{jm.hu@siat.ac.cn}
% \affiliation{%

\author{Jianmin Hu}
%\orcid{0000-0000-0000-0000}
\affiliation{%
  \institution{Southern University of Science and Technology; Shenzhen Institutes of Advanced Technology, CAS}
  \streetaddress{Chinese Academy of Sciences}
  \city{Shenzhen}
  \state{Guangdong}
  \postcode{518055}
  \country{China}
  }

\author{Minxian Xu}
\authornote{Corresponding author}
\orcid{0000-0000-0000-0000}
\affiliation{%
  \institution{Shenzhen Institutes of Advanced Technology, CAS}
  \streetaddress{Chinese Academy of Sciences}
  \city{Shenzhen}
  \state{Guangdong}
  \postcode{518055}
  \country{China}
}

\author{Sa Wang}
\affiliation{%
  \institution{Institute of Computing Technology, CAS}
  \streetaddress{}
  \city{Beijing}
%  \state{Guangdong}
 % \postcode{518055}
  \country{China}
}

\author{Chong Ma}
\affiliation{%
  \institution{Alibaba Group}
  \streetaddress{}
  \city{Hangzhou}
%  \state{Guangdong}
 % \postcode{518055}
  \country{China}
}

\author{Min Shen}
\affiliation{%
  \institution{Alibaba Group}
  \streetaddress{}
  \city{Hangzhou}
%  \state{Guangdong}
 % \postcode{518055}
  \country{China}
}

\author{Kejiang Ye}
%\authornote{Corresponding author}
%\orcid{0000-0000-0000-0000}
\affiliation{%
  \institution{Shenzhen Institutes of Advanced Technology, CAS}
  \streetaddress{Chinese Academy of Sciences}
  \city{Shenzhen}
  \state{Guangdong}
  \postcode{518055}
  \country{China}
}

\author{Lin Qu}
\affiliation{%
  \institution{Alibaba Group}
  \streetaddress{}
  \city{Hangzhou}
%  \state{Guangdong}
 % \postcode{518055}
  \country{China}
}

\author{Chengzhong Xu}
%\authornote{Corresponding author}
%\orcid{0000-0000-0000-0000}
\affiliation{%
  \institution{University of Macau}
  \streetaddress{}
  \city{Macau}
  \state{Macau}
%  \postcode{518055}
  \country{China}
}

%\author{
%  Jianmin Hu$^{1,2}$, 
%  Minxian Xu$^2$, 
%  Sa Wang$^3$, 
%  Chong Ma$^4$, 
%  Min Shen$^4$, 
%  Kejiang Ye$^2$, \\
%  Lin Qu$^4$, 
%  Chengzhong Xu$^5$
%}

% ==================== 统一的机构列表 ====================
%\affiliation{
%  \institution{
%    $^1$Shenzhen Institutes of Advanced Technology, Chinese Academy of Sciences \quad
%    $^2$Southern University of Science and Technology \quad
 %   $^3$Institute of Computing Technology, Chinese Academy of Sciences \\
%    $^4$Alibaba Group \quad
%    $^5$University of Macau
%  }
%  \country{} % 必须保留这行留空，以防ACM模板报错
%}


\begin{abstract}
Multi-turn conversation is a fundamental scenario in LLM applications, widely used in chatbots and AI agents. As the conversation evolves, historical tokens accumulate continuously. Existing systems cache their key-value (KV) pairs to avoid redundant computation. However, limited GPU memory (HBM) capacity often forces these KV caches to be offloaded to CPU memory or SSD, making KV cache reloads increasingly costly in terms of latency as the context grows. Meanwhile, the constrained HBM capacity also limits the maximum inference length, thereby restricting the number of turns that can be supported in a conversation.

To address these two challenges, we propose SwiftCache, a collaborative inference system that enables heterogeneous models to share underutilized GPU memory and NVLink bandwidth within a server. Specifically, models with low KV cache demand donate idle GPU memory to store the prefix cache of high-demand models, allowing cross-model KV cache sharing over NVLink and avoiding slow PCIe transfers. SwiftCache further reduces memory pressure by keeping only the KV cache of the currently active layer in local GPU memory, thereby enabling longer-context inference.
Our experiments on real-world workloads show that SwiftCache reduces P99 time-to-first-token (TTFT) by up to 69\% and extends maximum context length by up to 3.98$\times$ compared to vLLM and SGLang, with minimal interference to co-located models.
\end{abstract}

\begin{CCSXML}
<ccs2012>
 <concept>
  <concept_id>00000000.0000000.0000000</concept_id>
  <concept_desc>Do Not Use This Code, Generate the Correct Terms for Your Paper</concept_desc>
  <concept_significance>500</concept_significance>
 </concept>
 <concept>
  <concept_id>00000000.00000000.00000000</concept_id>
  <concept_desc>Do Not Use This Code, Generate the Correct Terms for Your Paper</concept_desc>
  <concept_significance>300</concept_significance>
 </concept>
 <concept>
  <concept_id>00000000.00000000.00000000</concept_id>
  <concept_desc>Do Not Use This Code, Generate the Correct Terms for Your Paper</concept_desc>
  <concept_significance>100</concept_significance>
 </concept>
 <concept>
  <concept_id>00000000.00000000.00000000</concept_id>
  <concept_desc>Do Not Use This Code, Generate the Correct Terms for Your Paper</concept_desc>
  <concept_significance>100</concept_significance>
 </concept>
</ccs2012>
\end{CCSXML}

\ccsdesc[500]{Computer systems organization~Real-time systems}
\ccsdesc[300]{Computing methodologies~Natural language processing; Distributed computing methodologies}
% \ccsdesc[100]{Distributed computing methodologies.}
% \ccsdesc[100]{Do Not Use This Code~Generate the Correct Terms for Your Paper11}

\keywords{Large Language Models, KV Cache, Multi-Model Serving, Cross-Model Memory Sharing}

\maketitle

\section{Introduction}

 With the rapid advancement of large language models (LLMs), the maximum context window supported by modern LLMs has increased dramatically. Many state-of-the-art LLMs now support inference with context windows ranging from 128K to 1M tokens. Multi-turn conversation is a critical and highly prevalent long-context inference scenario, widely used in chatbots~\cite{openaiChatGPT,qwen,deepseek,claude}, AI agents~\cite{autogpt,openai_agents_python}, and other interactive applications. In multi-turn conversations, the amount of historical tokens increases as the number of turns grows, because the content from all previous turns, primarily consisting of user prompts and system responses, must be preserved and continuously incorporated into inference. In contrast, each new turn typically contributes only a relatively small amount of additional user input or response content. The KV pairs corresponding to historical tokens are usually cached in HBM. However, as the number of historical tokens increases, HBM eventually becomes insufficient to accommodate additional KV pairs. As a result, part of the KV pairs must be offloaded to CPU memory or SSD. Such a hierarchical caching strategy has already been adopted by mainstream LLM serving systems, including vLLM~\cite{kwon2023efficient,liu2025lmcache}, SGLang~\cite{zheng2024sglang}, and Mooncake~\cite{qin2025mooncake}. However, existing hierarchical caching mechanisms still face two major challenges when serving long-context multi-turn conversations.

\begin{figure}[tb]
  \centering
  \includegraphics[width=\columnwidth]{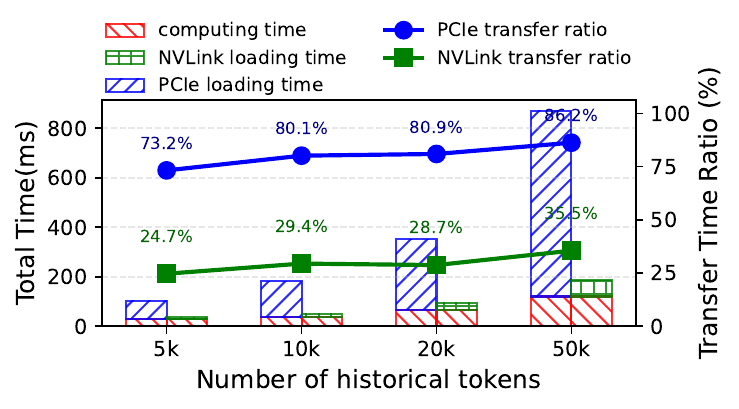}
    \caption{Latency breakdown for computing 256 new tokens and loading KV cache with varying numbers of historical tokens on two 96\,GB H20 GPUs using the LWM-1M-Text~\cite{liu2025world}. The system uses PCIe~4.0 with a peak bandwidth of 32\,GB/s and NVLink with a peak bandwidth of 400\,GB/s.}
  \label{fig:ratio}
  \vspace{-0.5em}
\end{figure}

\textbf{First, KV cache reload becomes increasingly expensive as history grows.} As conversational context continues to grow, TTFT rises substantially even though newly entered input remains relatively stable across turns~\cite{gao2024cost}. This latency growth is mainly due to the cost of fetching historical KV caches from CPU memory or SSD back into GPU memory over PCIe, which becomes the dominant bottleneck. As shown in Figure~\ref{fig:ratio}, PCIe-based loading takes around 75ms at 5k historical tokens, while computation only takes about 27ms, and the transfer overhead rises from 73.2\% to 86.2\% as history grows to 50k tokens. This suggests that in multi-turn conversations, KV cache transfer for historical tokens is non-negligible and can dominate the overall latency. By contrast, NVLink can substantially reduce this transfer overhead and mitigate its impact on end-to-end inference latency.

\textbf{Second, limited HBM capacity restricts the maximum context length.} Although many modern LLMs advertise very long context windows, the actual length they can handle is often much smaller due to HBM limits. For example, although LWM-1M-Text~\cite{liu2025world} supports a maximum context length of 1M tokens, it can only process sequences up to 18K tokens on a 32GB V100 GPU and up to 160K tokens even on a 96GB H20 GPU. Existing approaches such as chunked prefill~\cite{agrawal2024taming}, parallelization~\cite{shoeybi2019megatron,lepikhin2020gshard,huang2019gpipe}, and compression-based methods~\cite{lee2024infinigen,zhang2023h2o,zhang2025diffkv,hu2026brownoutserve} can help, but they either require more hardware or may hurt accuracy. Meanwhile, we observed that some co-located models still leave idle GPU memory, creating an opportunity to extend inference length by sharing memory across models.

To address these two challenges, we propose SwiftCache, an efficient inference system designed for long-context multi-turn conversations. The core ideas of SwiftCache are as follows: \textbf{(1) it offloads KV pairs into idle blocks in the KV caches of other models via high-speed NVLink to accelerate KV cache transmission.} This is motivated by the observation that heterogeneous models co-located on the same server often serve workloads with markedly different KV cache demands: some models handling long-context multi-turn conversations require large cache capacity, while others serving shorter or less cache-intensive workloads leave substantial KV cache space idle. Meanwhile, because different models usually run independently, they rarely generate frequent GPU-to-GPU communication with each other at the hardware level. SwiftCache leverages this otherwise idle bandwidth and memory for fast cross-model KV cache sharing. \textbf{(2) it retains only the KV pairs of the currently active layer in the local KV cache, thereby reducing GPU memory overhead and supporting longer inference lengths.} We observed that only the currently executing layer accesses its own KV cache, while the KV caches of other layers remain temporarily idle but still occupy valuable GPU memory. Existing systems~\cite{kwon2023efficient, zheng2024sglang, lightllm} typically keep all layer KV caches resident in local GPU memory, but this design significantly increases peak memory usage and limits the maximum context length. \textbf{(3) it introduces an elastic cache mechanism that can be resized in \(\mathbf{O(1)}\) time according to workload intensity.} In multi-model serving, cache capacity needs to be flexibly reallocated as workload intensity changes over time. However, existing systems often incur substantial data movement if KV cache resizing is required because of the \textit{layer-major} layout. To tackle this, SwiftCache proposes a \textit{block-major} layout and integrates it into the elastic cache, enabling fast, low-overhead reallocation.

Overall, SwiftCache is designed as a multi-model collaborative inference system for long-context workloads. It enables KV cache memory space sharing across co-located models, allowing models with high KV cache demand to offload prefix KVs to idle cache blocks of other models. The core components of SwiftCache are \textit{layer stream cache} and \textit{elastic cache}. Layer stream cache stores only the KV pairs of the currently active layer in local GPU memory. It streams the KV pairs of other layers from external memory on demand, thereby reducing memory pressure and extending the maximum inference length (\S\ref{sec:master_cache_manager}). To further support flexible and efficient cache reallocation, SwiftCache introduces elastic cache with \textit{block-major} layout, which dynamically adjusts the cache capacity assigned to each model with only \(O(1)\) overhead according to workload demand (\S\ref{sec:worker_cache_manager}). Together, these designs improve GPU memory utilization and overall serving performance for long-context inference.

We evaluate SwiftCache on two real-world multi-turn conversation datasets, ShareGPT~\cite{sharegpt90k} and L-Eval~\cite{an2024eval}, using Qwen3-family~\cite{yang2025qwen3} models and LWM-1M-Text~\cite{liu2025world}. Compared with state-of-the-art inference systems, SwiftCache reduces P99 TTFT by up to 69\% and extends the maximum inference length by up to $3.98\times$ across GPUs with various HBM capacities.
Our key \textbf{contributions} are:
\begin{itemize}[itemsep=0.5em]
    \item We analyze the characteristics of multi-turn conversations and identify two key challenges for LLM serving in this scenario: inefficient KV cache transfer and limited HBM capacity.

    \item We propose SwiftCache, a collaborative serving system that enables heterogeneous models to share KV cache through NVLink and dynamically reallocate memory across models, fully utilizing underused NVLink bandwidth and GPU memory within a server.
    
    \item We design layer stream cache and elastic cache, two mechanisms that reduce local KV footprint and support efficient \(O(1)\) resizing of cache allocations.
    
    \item We demonstrate significant gains in latency and maximum context length across real-world workloads and different GPU memory capacities, with only minimal interference to co-located models.
\end{itemize}
% \vspace{1em}
\section{Background and Motivations}~\label{sec:background_motivation}
% \vspace{-1.5em}
\subsection{KV Cache and Prefix Caching}
Autoregressive LLMs produce tokens one after another, and each token looks back at all the tokens generated before it. In each iteration, the model recomputes the keys and values for all preceding tokens in the sequence. To reduce this redundancy, modern inference systems~\cite{kwon2023efficient,zheng2024sglang,lightllm} widely adopt KV cache, which stores the key and value states of past tokens for reuse in later decoding steps. By reusing these cached states, KV cache significantly improves decoding efficiency and is a core mechanism in LLM inference.

In addition to intra-sequence reuse during generation, many workloads also exhibit prefix reuse across different requests, where multiple prompts share the same system instruction, template, or other common prefix. KV cache can store the KV pairs from a processed request. When a subsequent request shares the same prefix, these cached pairs can be reused to avoid redundant computation. We refer to this cross-request reuse of KV cache as \textit{prefix caching}~\cite{gao2024cost,zhang2025jenga,lightllm}. One typical use case of prefix caching is multi-turn conversations. Since LLMs do not have an intrinsic memory mechanism, multi-turn interaction is implemented as a sequence of independent requests, where the prompt of each turn is constructed by concatenating the previous content with the user’s current input. Consequently, all requests in a session share the same prefix, which makes multi-turn conversations a natural use case for prefix caching.
\begin{figure}[tb]
  \centering
  \includegraphics[width=\columnwidth]{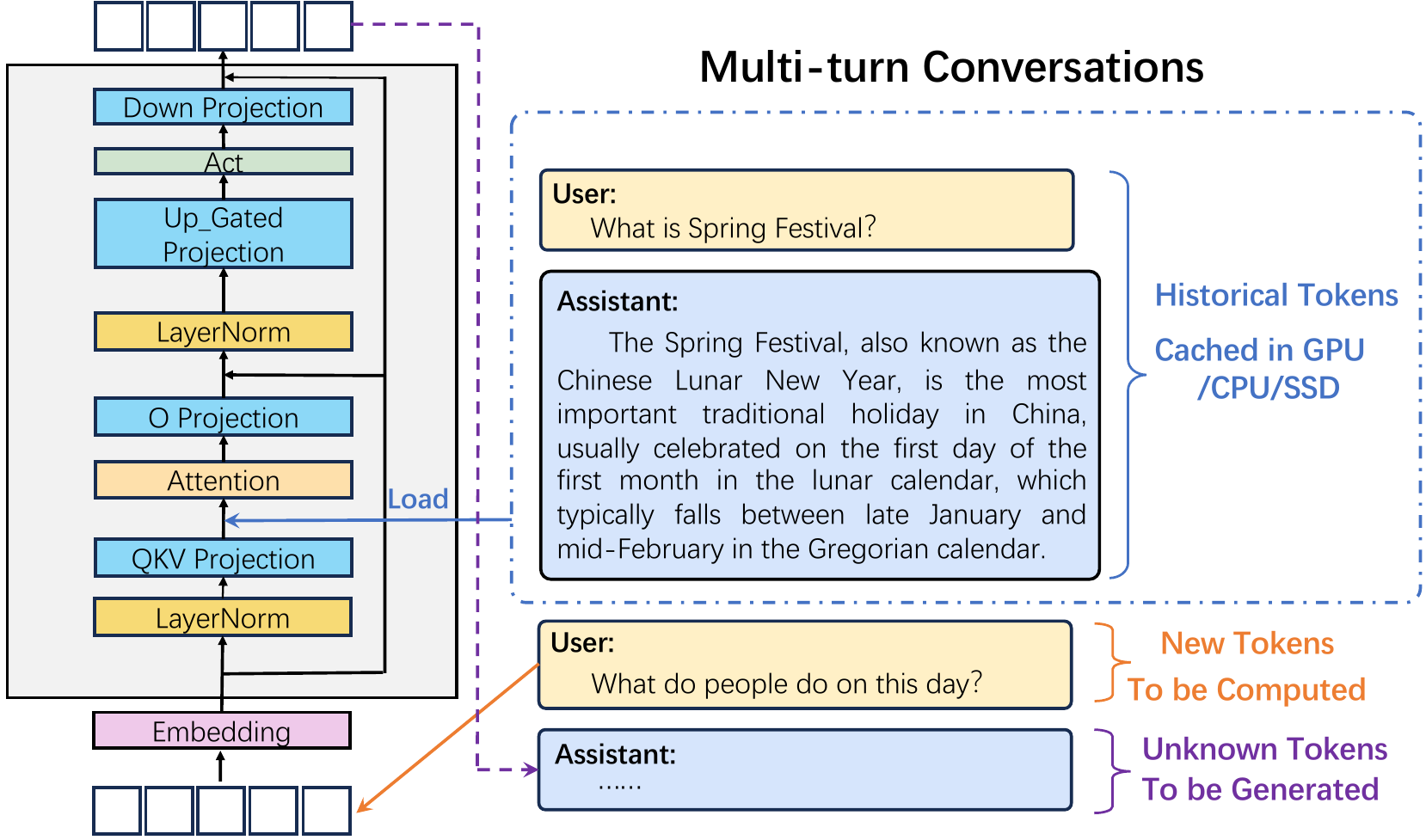}
    \caption{An illustration of multi-turn conversation inference in LLMs. Historical tokens from previous turns are cached in GPU, CPU, or SSD memory, while only new tokens in the current turn need to be computed. During attention, the cached key-value pairs of historical tokens are loaded and combined with newly computed tokens to generate the next response.}
  \label{fig:mult_turn_conversations}
  \vspace{-0.5em}
\end{figure}

\subsection{Multi-turn Conversations Inference in LLMs}
\textbf{Transformer architecture.} Transformer~\cite{vaswani2017attention} is the backbone architecture of modern LLMs. As shown in Fig.~\ref{fig:mult_turn_conversations}, a transformer layer typically consists of embedding, self-attention, feed-forward network, and normalization modules. Given an input sequence, the model first maps tokens into hidden representations through the embedding layer. These representations are then projected into query, key, and value vectors via the QKV projection. In the attention module, each token attends to all tokens that precede it to capture contextual dependencies. The resulting representations are further processed by output projection and feed-forward layers, which usually include gated-up projection, activation, and down-projection operations. By stacking multiple Transformer layers, LLMs can model complex semantic relationships over long sequences.
\begin{figure}[t]
  \centering
  \begin{subfigure}[t]{0.48\columnwidth}
    \centering
    \includegraphics[width=\linewidth]{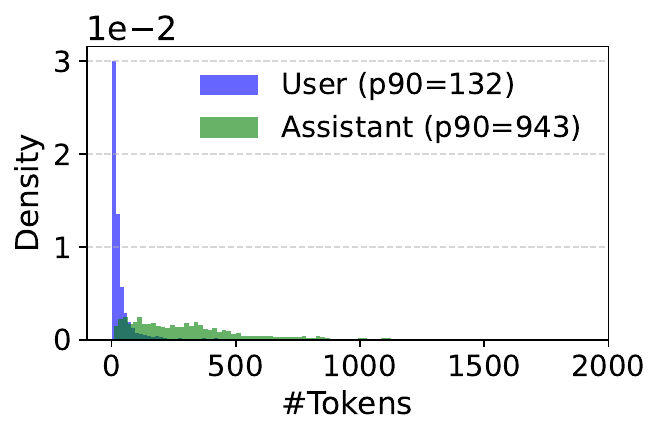}
    \caption{Distributions of user prompt and assistant response length.}
    \label{fig:human_gpt_token_distribution}
  \end{subfigure}
  \hfill
  \begin{subfigure}[t]{0.48\columnwidth}
    \centering
    \includegraphics[width=\linewidth]{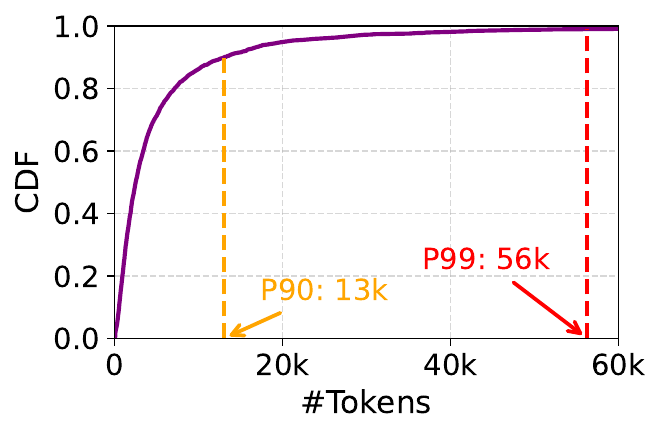}
    \caption{Distribution of total token lengths across sessions.}
    \label{fig:multi_turn_dialog_token_sum_cdf}
  \end{subfigure}
    \caption{Token length statistics in ShareGPT.}
  \label{fig:mainfig}
\end{figure}

\begin{table*}[t]
\centering
\small
\setlength{\tabcolsep}{4pt}
\renewcommand{\arraystretch}{1.1}
\caption{Performance of the LWM-1M-Text model across various workloads on an H20 GPU with and without prefix cache.}
\label{table:workloads_characters}
\begin{tabular}{c|c|c|c|c|c|c}
\toprule
\textbf{Workload Classifications} & \textbf{Data Source} & \textbf{Hit Rate (\%)} & \textbf{TTFT w/ Cache} & \textbf{TTFT w/o Cache} & \textbf{Avg. Input Length} & \textbf{Demand} \\
\midrule
Summarization & L-Eval~\cite{an2024eval} & 6.3  & 0.77s & 0.86s & 7124 & Low \\
Code Completion & HumanEval~\cite{chen2021evaluating} & 0.1  & 0.03s & 0.03s & 155 & Low \\
Multi-turn Conversations & ShareGPT~\cite{sharegpt90k} & 90.7 & 0.11s & 0.72s & 6231 & High \\
Question Answering & L-Eval~\cite{an2024eval} & 81.8 & 0.92s & 7.95s & 43958 & High \\
\bottomrule
\end{tabular}
\end{table*}
\noindent \textbf{Multi-turn conversations.} Multi-turn conversations are a common interaction pattern for LLMs, where a user and the assistant exchange multiple rounds of messages. In this setting, the tokens from previous turns form the conversational history, while the newly entered user tokens constitute the current input. During inference, only the new tokens need to be computed, whereas the historical tokens can be reused through the KV cache. Compared with recomputing all historical tokens from scratch, using the KV cache avoids repetitive execution of embedding, feed-forward network, normalization, and projection operations, which are essentially large matrix multiplications, thereby reducing the overall computational cost of inference. Specifically, in the attention computation of a new turn, the model retrieves the cached KV pairs of historical tokens and combines them with the KV pairs from the new tokens. This mechanism allows the model to avoid recomputing previously seen tokens, significantly reducing the TTFT of the next turn. When the historical tokens’ KV pairs can be reused in the next round, we refer to this as a \emph{prefix cache hit}.

\noindent \textbf{Characteristics of multi-turn conversations.} In multi-turn conversations, historical tokens accumulate across turns, whereas each new turn typically adds only a short user input. Thus, inference latency is driven not only by new-token computation but also by historical KV cache loading, which becomes expensive when cache is offloaded to CPU memory or SSD. As shown in Fig.~\ref{fig:human_gpt_token_distribution}, 90\% of user prompts are shorter than 132 tokens, indicating that prefill computation is relatively lightweight. In contrast, Fig.~\ref{fig:multi_turn_dialog_token_sum_cdf} shows that 10\% and 1\% of sessions exceed 13k and 56k tokens, respectively, leading to severe GPU memory pressure and limiting maximum inference length.

% \section{Motivation}\label{sec:motivation}
\subsection{Different Workloads Have Different Prefix Cache Demands}
In LLM inference, we observe that the demand for prefix cache varies substantially across workloads. As Table~\ref{table:workloads_characters} shows, multi-turn conversational and question answering scenarios often involve extensive reuse of prefix KV pairs, with prefix cache hit rates of 90.7\% and 81.8\%, respectively, making prefix caching a critical mechanism for reducing inference latency. Therefore, models serving such workloads require large KV cache capacity for higher prefix cache hit rates. In contrast, tasks like summarization or code completion typically present distinct contexts, resulting in low hit rates and minimal cache usage, with prefix cache hit rates of 6.3\% and 0.1\%, respectively. Turning off prefix cache in these tasks would not significantly impact performance. This difference in KV cache demand highlights the sharply different KV cache capacity requirements of LLM services across scenarios, and suggests an opportunity to optimize inference through cross-model KV cache sharing.

\vspace{0.5em}
\noindent \textit{\textbf{Insight-1}: Models with high KV cache capacity demand can offload their prefix cache to the KV cache space of lower demand models, thereby improving overall GPU memory utilization.}

\subsection{Underutilized NVLink Bandwidth Between Heterogeneous Models}
In a server, where each model may occupy one or more GPUs, we observe that GPUs belonging to different models rarely engage in direct GPU-to-GPU communication. This is because mainstream parallelization strategies, including Tensor Parallelism~\cite{shoeybi2019megatron}, Model Parallelism~\cite{li2023alpaserve}, and Pipeline Parallelism~\cite{huang2019gpipe}, operate exclusively within the scope of a single model, leaving high-bandwidth interconnects such as NVLink or NVSwitch between GPUs of different models largely underutilized. Even in multi-model collaborative scenarios such as multi-agent systems, inter-model communication is typically mediated by the CPU through network or process-level messaging rather than direct GPU-to-GPU transfer. As a result, NVLink links between heterogeneous models often remain idle or only lightly utilized. Therefore, if we co-locate a model with high KV cache demand and a model with low KV cache demand within a server, the former can offload its prefix cache to the GPUs of the latter via these high-speed interconnects. Compared to the conventional approach of offloading the prefix cache to host memory or SSD through PCIe, which incurs higher latency and CPU involvement, this direct GPU-to-GPU transfer is significantly faster and fully bypasses the CPU.

\vspace{0.5em}
\noindent \textit{\textbf{Insight-2}: We can co‑locate models with high KV cache capacity demand and models with low demand on a server, and transfer prefix cache via NVLink.}

\subsection{Only One Layer of KV Cache is Active During LLM Inference}
In LLM inference, the attention of each layer only accesses the KV cache of that layer. KV caches from other layers remain idle but still occupy GPU memory. This means that most of the KV cache is inactive while a given layer is running. Based on this observation, all layers’ KV caches can be stored in external memory such as another GPU, CPU memory, or high-speed devices. The local GPU only keeps a buffer for the current layer. When executing the attention of a certain layer, its KV cache is loaded from the external memory into the local GPU in advance. This way, the GPU only needs memory space for one layer’s KV cache at a time, allowing that space to store more KV pairs and support a longer context length.

For example, consider a model with three decoder layers on a given GPU, where each layer can store up to 10K tokens of KV cache. If the KV caches of the other two layers are stored on other GPUs, the local GPU can repurpose all KV cache memory, originally allocated for the KV pairs of all three layers, to store just a single layer, theoretically increasing its capacity to 30K tokens. This both preserves the KV caches of other layers and increases the maximum context length during inference.

Furthermore, based on \textit{\textbf{Insight-1}} and \textit{\textbf{Insight-2}}, when high KV cache capacity demand models and low demand models are co-located on the same machine, the prefix cache of the high demand model can be transferred to the GPUs of the low‑demand model via high-speed NVLink interconnects. This direct GPU‑to‑GPU migration avoids PCIe and CPU involvement, thereby preventing high inference latency from cache transfer. 

\vspace{0.5em}
\noindent \textit{\textbf{Insight-3}: Inactive KV caches from other layers in LLM inference can be offloaded, enabling the active layer to store more KV and extend the maximum context length.}
\section{SwiftCache Design}\label{sec:overview}
\begin{figure}[tb]
  \centering
  \includegraphics[width=\columnwidth]{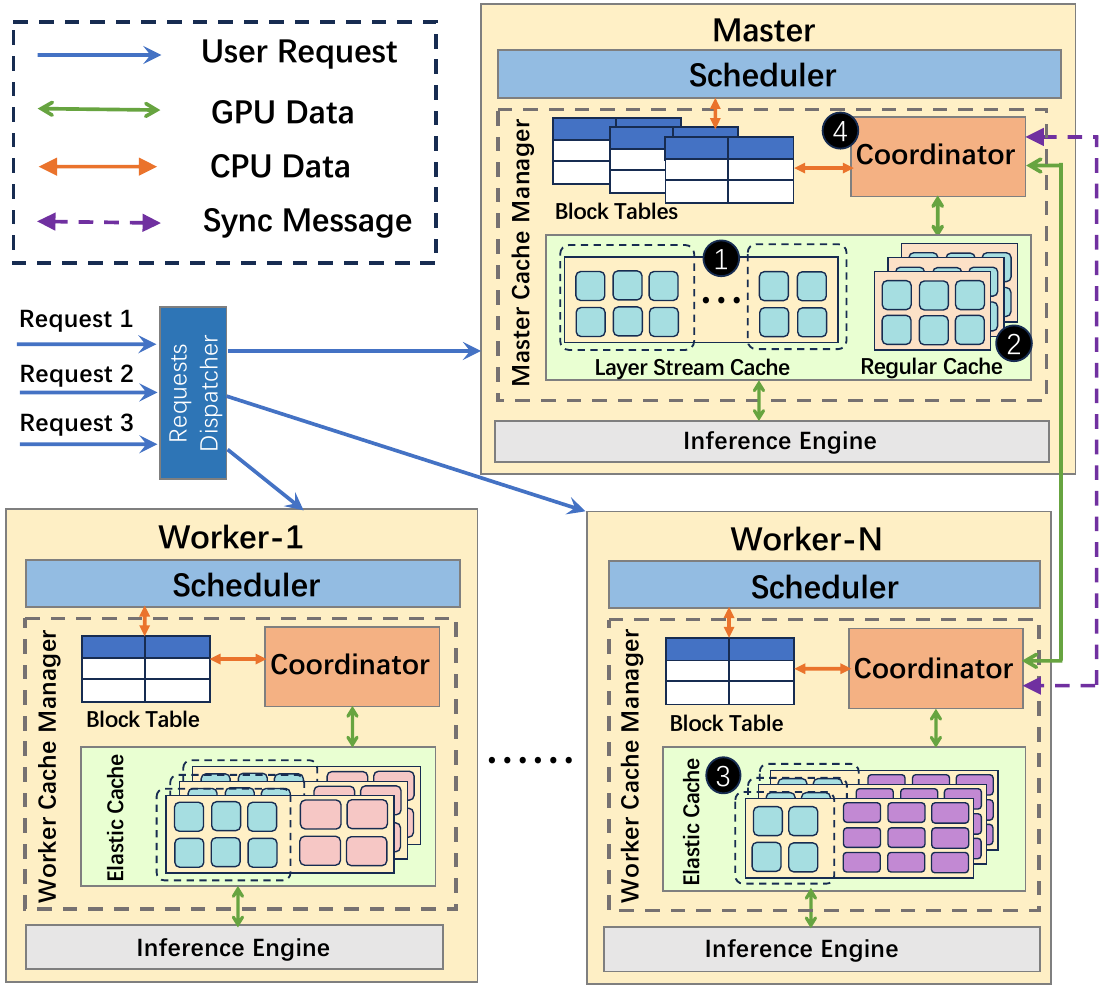}
  \caption{SwiftCache system overview.}
  \label{fig:system_overview}
  \vspace{-0.5em}
\end{figure}
\subsection{Overview}
To address the above challenges we propose SwiftCache, a multi-model collaborative system that is based on our three insights to improve GPU memory utilization and NVLink bandwidth efficiency within a server. As shown in Fig.~\ref{fig:system_overview}, SwiftCache consists of one model with high KV cache demand, referred to as the \textit{master}, and multiple models with low KV cache demand, referred to as \textit{workers}. Each model has its own \textit{scheduler}, \textit{cache manager}, \textit{coordinator}, and \textit{inference engine}. User requests are dispatched to either the master or a worker based on their KV cache demand, and the scheduler adopts an iteration-level~\cite{yu2022orca} scheduling policy with FCFS ordering.

On the master side, the master cache manager maintains two cache structures: layer stream cache and regular cache. The layer stream cache keeps only the KV cache of the active layer in local GPU memory to reduce local memory usage, while the regular cache stores full-layer KV blocks in the master's local memory for low-latency reuse, without incurring any cross-GPU data transfer        (\S~\ref{sec:master_cache_manager}). On the worker side, the worker cache manager maintains an elastic cache with a \textit{block-major} layout that supports $O(1)$ cache resizing between local workloads and master borrowing (\S~\ref{sec:worker_cache_manager}). When a worker is lightly loaded, it donates idle cache blocks to the master; when its own demand increases, it reclaims the borrowed capacity. This constant-time resizing is enabled by block-level remapping, allowing each worker cache manager to update its local block table and coordinate synchronization without expensive data movement (\S~\ref{sec:coordination}). The master and workers interact through their respective coordinators to ensure the elastic cache is reasonably allocated between them (\S~\ref{sec:coordination}).

\subsection{Master Cache Manager Design} \label{sec:master_cache_manager}
In this section, we describe the two key components of the master cache manager, layer stream cache and regular cache, and show how they together enable SwiftCache to extend the maximum inference length.

\par\medskip
\noindent\textbf{Layer Stream Cache.} 
% \label{sec:layer_stream_cache}
Layer stream cache (LSC, denoted by \blackcircled{1} in Figure~\ref{fig:system_overview}) is a GPU buffer designed to temporarily hold the KV cache of only the currently executed layer during inference. In scenarios where the requested prefix cache is stored on a worker GPU, SwiftCache preloads the KV cache for a given layer into the LSC of the master GPU before the model executes the attention of that layer. Newly generated KV entries are also written into this memory and subsequently transferred back to the worker GPU after computation. Because LSC stores only one layer’s KV cache at any given time, it significantly reduces the amount of KV cache that needs to reside in the master’s local GPU memory, thus enabling a longer maximum inference context length. 

The capacity of LSC, measured in terms of the number of blocks it can hold, is determined by the total KV cache size of all worker models and the memory footprint of a single block in the master model. Specifically, let the total KV cache size across all worker models be \(\sum_{i=1}^{N} C_{\text{worker}}^{(i)} \), where \(C_{\text{worker}}^{(i)}\) denotes the KV cache size of the \(i\)-th worker model and \(N\) is the total number of worker models. In the master model, the memory footprint of a single KV block per layer is given by:

\begin{equation}
M_{\text{block}} = 2 \times B \times H_{\text{kv}} \times D_{\text{kv}} \times d_{\text{type}},
\end{equation}
where the factor of 2 reflects the storage of both key and value tensors, \(B\), \(H_{\text{kv}}\), \(D_{\text{kv}}\), and \(d_{\text{type}}\) denote the block size, the number of KV heads, dimension per KV head, and the number of bytes per element in the master, respectively.

Let \(K_i\) denote the number of master full-layer blocks that can be stored using the KV cache space of the \(i-\)th worker model, i.e.,

\begin{equation}
K_i = \left\lfloor \frac{C_{\text{worker}}^{(i)}}{M_{\text{block}} \times L} \right\rfloor,
\end{equation}
where \(L\) denotes the number of decoder layers of the master.

Let \(K_{\text{master}}\) denote the number of single-layer blocks that the master's local memory can store:

\begin{equation}
K_{\text{master}} = \left\lfloor \frac{C_{\text{master}}}{M_{\text{block}}} \right\rfloor,
\end{equation}
where \(C_{\text{master}}\) is KV cache size of the master.

Then, the number of KV blocks in LSC is given by:

\begin{equation}
N_{\text{LSC}} = \min\left( \sum_{i=1}^{N} K_i,\; K_{\text{master}} \right).
\label{eq:n_lsc}
\end{equation}

This formulation ensures that LSC is sized to hold the KV cache from all worker devices, up to the master's local memory limit, and enables the master to process long sequences without exhausting its memory.

\par\medskip
\noindent\textbf{Regular Cache.}
~\label{sec:regular_cache}
Since the size of LSC is strictly determined by Eq.~\ref{eq:n_lsc}, the master's local GPU memory may not be fully occupied. When \(\sum_{i=1}^{N} K_i < K_{\text{master}}\), LSC consumes exactly \(\sum_{i=1}^{N} K_i\) single-layer blocks, leaving surplus memory for regular cache (RC, denoted by \blackcircled{2} in Figure~\ref{fig:system_overview}). RC stores KV cache directly on the master's local memory and retains full-layer data throughout the entire inference process, offering the lowest access latency as it requires no cross-GPU transfer. Using the notation defined above, the number of full-layer KV blocks that RC can store is given by:

\begin{equation}
N_{\text{RC}} = \left\lfloor \frac{K_{\text{master}} - \sum_{i=1}^{N} K_i}{L} \right\rfloor.
\label{eq:n_rc}
\end{equation}

Conversely, when \(\sum_{i=1}^{N} K_i \geq K_{\text{master}}\), LSC occupies the entire master memory, leaving no space for RC. Ultimately, the maximum inference length supported by the master is \((N_{\text{LSC}} + N_{\text{RC}}) \times B\) tokens.

Here is a specific example. Assume the master has \(L = 10\) layers. The master's local memory can store a total of \(K_{\text{master}} = 100\) single-layer blocks. Worker 1 can store \(K_1 = 9\) full-layer blocks, and worker 2 can store \(K_2 = 8\) full-layer blocks. The number of blocks that LSC can store is \(N_{\text{LSC}} = 9+8 = 17\). Thus, LSC occupies 17 blocks, leaving \(100 - 17 = 83\) single-layer blocks in master memory for RC. Since RC stores full-layer blocks, the number of RC blocks is \(N_{\text{RC}} = \left\lfloor 83/10 \right\rfloor = 8\).
Therefore, the maximum inference context length supported is \(N_{\text{LSC}} + N_{\text{RC}} = 25\) blocks. In contrast, a conventional system without LSC can only support \(\lfloor 100 / 10 \rfloor = 10\) blocks.

\subsection{Overlapping KV Cache Transfer with Computation}
\label{sec:overlapping}
To minimize the latency introduced by cross-GPU KV cache transfers, both load and store operations are performed in a layer-wise manner and can be overlapped with computation.

\par\medskip
\noindent \textbf{KV Store.}
Newly generated KV entries are produced after the QKV operation of each layer. Once generated, they can be asynchronously transferred to the worker GPU without blocking subsequent computation.

\par\medskip
\noindent \textbf{KV Load.}
For the first layer, the KV cache loading can begin immediately after the scheduling phase, before the attention computation starts. For subsequent layers (i.e., layer \(i\) where \(i \geq 2\)), the KV cache of layer \(i\) must wait until both the attention of layer \(i-1\) has completed and the layer-wise KV cache transfer of layer \(i-1\) has finished. This ensures that no data is overwritten before being successfully transferred back to the worker GPU. Once these conditions are satisfied, the KV cache of layer \(i\) can be loaded asynchronously while the FFN of layer \(i-1\) executes, effectively hiding the transfer latency.

By leveraging layer-wise pipelining, overlapping computation with communication, and the high-bandwidth NVLink interconnect, the overhead of both loading from and storing to worker GPUs can be effectively hidden, thereby improving overall inference throughput.

\subsection{Worker Cache Manager Design} ~\label{sec:worker_cache_manager}
\textbf{Elastic Cache.}
% ~\label{sec:elastic_cache}
In multi-model inference scenarios, worker models must continue serving their own requests concurrently with the master model. Consequently, it is infeasible for a worker model to permanently allocate all of its GPU memory reserved for KV cache to the master. To address this constraint, we propose the elastic cache (denoted by \blackcircled{3} in Figure~\ref{fig:system_overview}) that enables the worker to dynamically reallocate portions of its KV cache memory to the master according to workload intensity.

% \subsection{How can Elasticity be Achieved Efficiently?}\label{sec:elasticity}

% \begin{figure}[tbp]
%   \centering
%   \begin{minipage}{0.19\textwidth}
%     \centering
%     \includegraphics[width=\textwidth]{figs/layer_major_scale_down.png}
%     \subcaption{scale down}
%     \label{fig:scale_down}
%   \end{minipage} \hfill
%   \begin{minipage}{0.23\textwidth}
%     \centering
%     \includegraphics[width=\textwidth]{figs/layer_major_scale_up.png}
%     \subcaption{scale up}
%     \label{fig:scale_up}
%   \end{minipage}
%   \caption{Resizing operations in a \emph{layer-major} KV cache layout.}
%     \label{fig:resizing operations}
%     \vspace{-1em}
% \end{figure}

\begin{figure}[tb]
  \centering
  \includegraphics[width=\columnwidth]{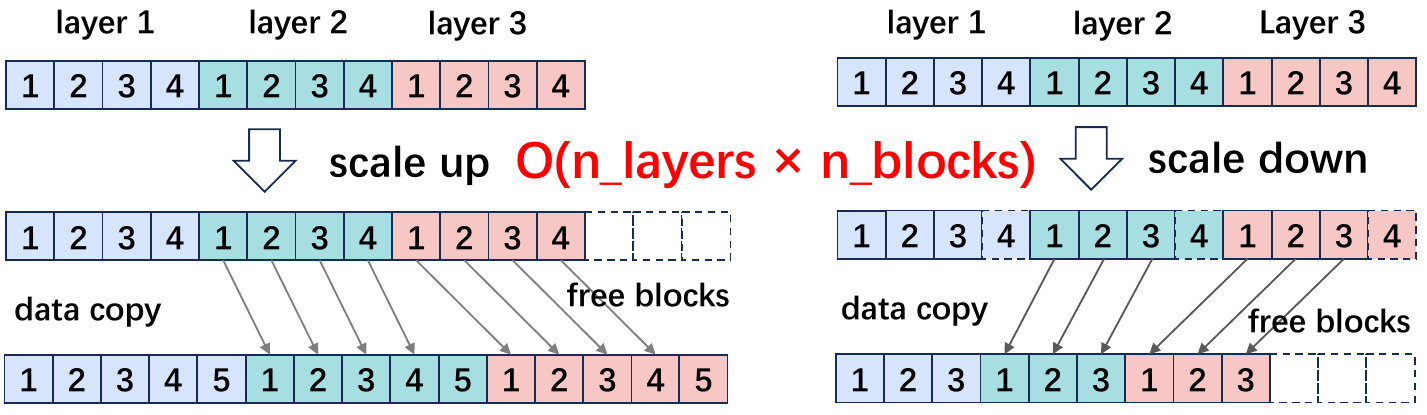}
  \caption{Resizing operations in a layer-major KV cache layout. }
  \label{fig:layer-major}
  \vspace{-0.5em}
\end{figure}

\begin{figure}[tb]
  \centering
  \includegraphics[width=\columnwidth]{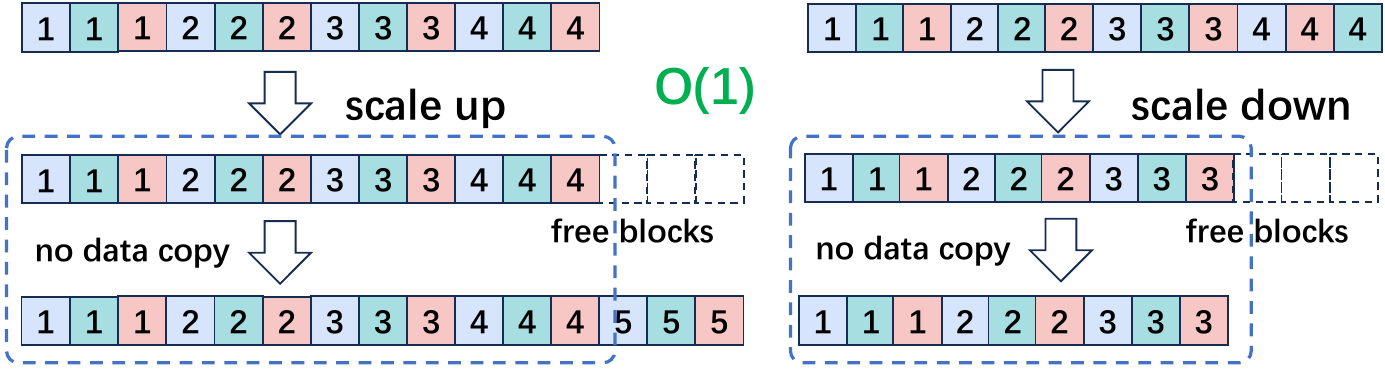}
  \caption{Resizing operations in a block-major KV cache layout. }
  \label{fig:block-major}
  \vspace{-0.5em}
\end{figure}

\par\medskip
\noindent \textbf{High Cost of Elasticity in Existing Systems}. Most current inference frameworks (e.g., vLLM~\cite{kwon2023efficient}, SGLang~\cite{zheng2024sglang}, LightLLM~\cite{chen2025pre3,gong2025past}) are not designed for dynamic KV cache resizing. If such resizing operations are performed within these systems, they would incur substantial overhead due to their \emph{layer-major} layout. In this layout, memory is organized with the layer index as the first dimension and the block index as the second dimension, 
so the shape of the KV cache tensor is
(\texttt{n\_layers},\ \texttt{n\_blocks},\ \texttt{block\_elems}),
where
   \texttt{n\_layers} is the total number of transformer layers in the model.
   \texttt{n\_blocks} is the number of KV cache blocks allocated per layer.
   \texttt{block\_elems} is the number of elements in a block, computed as 
   \(\texttt{block\_size} \times \texttt{n\_heads} \times \texttt{head\_dim}\).
Under this layout, all KV blocks belonging to the same layer are stored contiguously in memory, 
while blocks from different layers reside in separate, non-adjacent regions.

When resizing, such as reducing the number of blocks allocated to a worker model or increasing the allocation for a master model, the change is applied uniformly across all layers. This means that the same block index must be added or removed in every layer. Under the layer-major layout, all blocks within a single layer are stored contiguously, but blocks with the same index from different layers reside in separate regions of GPU memory. As a result, removing a block from the worker requires shifting subsequent blocks within each affected layer to close the gap and retain that layer’s internal contiguous arrangement.  Similarly, inserting new blocks for the master entails moving existing blocks within each layer to make room for the additions. This per-layer rearrangement, although logically concerning only a single block index, induces extensive physical memory movement across multiple disjoint regions, leading to significant latency and reallocation overhead.

The left part of Figure~\ref{fig:layer-major} shows the scaling-up process of a model with three decoder layers, each containing four blocks. 
In this case, each layer is extended with one additional block, which causes extensive data movement. For instance, the first block of the second layer must give up its position to the newly inserted fifth block in the first layer, and then be shifted to occupy the position of the second block in the second layer, triggering similar moves for subsequent blocks. The right part of Figure~\ref{fig:layer-major} shows the scaling-down process of the same model, where the fourth block of each layer is freed. 
Because blocks within the same layer are stored contiguously and blocks from different layers reside in separate memory regions, removing the last block requires shifting all blocks starting from the first block of the second layer forward. For example, the first block of the second layer is moved to the position originally occupied by the fourth block of the first layer. Since blocks in the first layer remain in place while blocks in the other layers must be moved, both operations result in a time complexity of \(O(\texttt{n\_layers} \times \texttt{n\_blocks})\).\newline

\noindent \textbf{Block-Major Layout for Efficient Resizing.} To eliminate cross-layer migration, we propose a \emph{block-major} layout in which all layer data belonging to the same block are stored contiguously in device memory. The KV cache shape is adjusted to (\texttt{n\_blocks},\ \texttt{n\_layers},\ \texttt{block\_elems}). As a result, resizing operations require only appending a new block at the end for expansion or releasing the last block for contraction, without relocating data from other blocks. The complexity is thereby reduced from \(O(\texttt{n\_layers} \times \texttt{n\_blocks})\) to \(O(1)\).

In this layout, all KV cache blocks with the same index across layers are stored contiguously in memory, while blocks of different indices are stored in separate regions. The left part of Figure~\ref{fig:block-major} shows the scaling-up process, where each layer is extended with one additional block (block index~5). 
Since the newly added blocks are placed at the end of their respective contiguous regions, no existing KV cache data needs to be moved.  The right part shows the scaling-down process, where the last block of each layer (block index~4) is freed. 
Removing these blocks simply releases the ending memory region without shifting any other blocks. Because both operations avoid moving existing KV cache data, the time complexity is \(O(1)\), making block-major layout highly efficient for dynamic resizing.

\par\medskip
\noindent \textbf{The Minimum Granularity of Scaling.} Due to differences in the number of layers, the size of each block, the number of attention heads, and the dimensionality of each head between the master and worker models, 
the number of elements contained in each KV cache block (\texttt{block\_elems}) also varies.  If KV cache borrowing and returning are performed at the granularity of a block, GPU memory alignment may not be preserved, leading to memory wastage.
Therefore, it is necessary to establish a shared alignment unit for memory scaling. 
This unit must ensure that, from both the master and worker perspectives, any expansion or shrinkage is performed in increments that are integer multiples of both KV cache block sizes. By doing so, memory waste can be avoided. 
This requirement naturally leads to the adoption of the \emph{least common multiple} (LCM)~\cite{zhang2025jenga} principle for determining the scaling unit.

We define the number of elements in a single block of the master model as
\begin{equation}
BE_{m} = L \times B \times H_{\text{kv}} \times D_{\text{kv}},
\end{equation}
and the number of elements in a single block of the \(i\)-th worker model as
\begin{equation}
BE_{i} = L_{i} \times B_{i} \times H_{\text{kv}}^{(i)} \times D_{\text{kv}}^{(i)},
\end{equation}
where \(L_{i}\), \(B_{i}\), \(H_{\text{kv}}^{(i)}\) and \(D_{\text{kv}}^{(i)}\) denote the number of layers, the block size, the number of KV heads and dimension per KV head, respectively.  The least common multiple of these quantities is denoted as
\begin{equation}
LCM_{\text{elem}} = \mathrm{lcm}(BE_{m}, BE_{i}),
\end{equation}
which serves as the shared alignment unit in terms of elements.  
From each model’s perspective, the minimum elastic unit (MEU) is then given by
\begin{equation}
MEU_{\text{m}} = \frac{LCM_{\text{elem}}}{BE_{m}}, \quad
MEU_{i} = \frac{LCM_{\text{elem}}}{BE_{i}}.
\end{equation}
This implies that whenever the master model scales up or down by \(MEU_{\text{m}}\) blocks, the \(i\)-th worker model must simultaneously scale down or up by \(MEU_{i}\) blocks.  
Furthermore, any scaling operation involving both the master and the \(i\)-th worker will necessarily touch at least \(MEU_{\text{m}}\) and \(MEU_{i}\) blocks, respectively.  

This strategy ensures that scaling operations are performed in units that are integer multiples of both the master and worker KV cache block sizes, thereby preserving GPU memory alignment and avoiding wastage.

\begin{algorithm}[t]
\caption{Elastic KV Cache Scale-Up and Scale-Down}
\label{alg:elastic_kv_cache}

\begin{algorithmic}[1]

\State \textbf{Input:}
\State \quad $N_i$ : Current \#KV blocks for $i$-th worker
\State \quad $B_i$ : Block size for $i$-th worker
\State \quad $MEU_i$ : Minimum elastic unit for $i$-th worker
\State \quad $MEU_m$ : Minimum elastic unit for master
\State \quad{$request\_len$: Number of tokens in the current request.}
\State \quad{$t$: Time window}
\State \textbf{Output:} Number of blocks needed to update for master and $i$-th worker
\State

\Function{ScaleUp}{$N_i, B_i, MEU_i, MEU_m, request\_len$}
    \State $need \gets \lceil request\_len / B_i \rceil$ \label{line:need_blocks}
    \If{$need \le N_i$} \label{line:check_satisfied_start}
        \State \Return $(0, 0)$ \label{line:check_satisfied_end}
    \Else
        \State $diff \gets need - N_i$ \label{line:calculate_diff_start}
        \State $k \gets \lceil diff / MEU_i \rceil$
        \State \Return $(k \times MEU_i,\ k \times MEU_m)$ \label{line:calculate_diff_end}
    \EndIf
\EndFunction
\State
\Function{ScaleDown}{$N_i, B_i, MEU_i, MEU_m, t$}
    \State $max\_len \gets \max\{len(r) \mid r \text{ in last } t \text{ seconds}\}$ \label{line:max_request_len_start}
    \State $max\_need \gets \lceil max\_len / B_i \rceil$ \label{line:max_request_len_end}
    \If{$max\_need \ge N_i$} \label{line:check_satisfied_start2}
        \State \Return $(0, 0)$ \label{line:check_satisfied_end2}
    \Else
        \State $diff \gets N_i - max\_need$ \label{line:calculate_diff_start2}
        \State $k \gets \lfloor diff / MEU_i \rfloor$
        \State \Return $(k \times MEU_i,\ k \times MEU_m)$ \label{line:calculate_diff_end2}
    \EndIf
\EndFunction

\end{algorithmic}
\end{algorithm}

\subsection{Master-Worker Coordination}~\label{sec:coordination} 
The master and worker models collaboratively manage KV cache capacity to balance serving demand and memory utilization. This cooperation is mediated by the coordinator (denoted by \blackcircled{4} in Figure~\ref{fig:system_overview}) in each cache manager, which handles cache borrowing, reclamation, and block-table synchronization. When the worker is under low load, its idle cache blocks can be borrowed by the master; when its own demand increases, it triggers a scale-up operation to reclaim capacity. This dynamic coordination enables SwiftCache to adapt cache allocation to changing workloads while preserving memory alignment and block mapping consistency.
% \vspace{-1em}
\par\medskip
\noindent \textbf{Worker KV Cache Scale-Up.} The worker scale-up operation is triggered when the current number of KV cache blocks allocated to a worker is insufficient to handle the incoming request. The scale-up procedure is described in the function \textsc{ScaleUp} in Algorithm~\ref{alg:elastic_kv_cache}. We first calculate the total number of KV cache blocks required for the incoming request (line~\ref{line:need_blocks}). If the required number of blocks is less than or equal to the worker’s currently allocated blocks, this indicates that the worker already has sufficient capacity to serve the request, and no scaling operation is performed (lines~\ref{line:check_satisfied_start}-\ref{line:check_satisfied_end}).

Otherwise, we determine the additional number of blocks needed to accommodate the request. This increment must be an integer multiple of the worker’s MEU to preserve memory alignment (lines~\ref{line:calculate_diff_start}-\ref{line:calculate_diff_end}). The worker cache manager then expands its local KV cache allocation accordingly and notifies the master cache manager through the coordinator. Upon receiving the synchronization message, the master cache manager updates its replica of the worker’s block table to reflect the new allocation. This synchronization ensures that both the worker and master maintain a consistent view of the block mapping after reallocation.
% \vspace{-0.5em}
\par\medskip
\noindent \textbf{Worker KV Cache Scale-Down.} The worker scale-down operation is performed periodically to determine whether the KV cache allocation for a worker can be reduced. The procedure is described in the function \textsc{ScaleDown} in Algorithm~\ref{alg:elastic_kv_cache}. At each check interval, the worker examines the longest request observed within the past time window $t$ (e.g., 1 minute) and uses it as an estimate for the potential length of future requests (lines~\ref{line:max_request_len_start}-\ref{line:max_request_len_end}). If the maximum number of KV cache blocks required by any request in this window is greater than or equal to the worker’s currently allocated blocks, we assume that similar requests may occur in the future, indicating that the current allocation is necessary. In this case, no scale down operation is performed (lines~\ref{line:check_satisfied_start2}-\ref{line:check_satisfied_end2}). Otherwise, when the maximum block requirement within the past time window is smaller than the worker’s current allocation, a scale-down is triggered. We calculate the difference between the current number of blocks and the past maximum requirement to determine the number of blocks to be released (lines~\ref{line:calculate_diff_start2}-\ref{line:calculate_diff_end2}). This number also must be an integer multiple of the worker's MEU. Similar to scale-up, the master and worker synchronize their block tables through their respective coordinators.

\section{Implementation}
SwiftCache is an end-to-end, high-performance inference system designed to optimize KV cache storage and transmission during the \emph{prefill} stage. It operates in single-server, multi-GPU environments to coordinate multiple heterogeneous models, fully utilizing both GPU memory and inter-GPU bandwidth resources. We employ FastAPI~\cite{fastapi} as the front-end interface for user requests. The inference engine components of SwiftCache are implemented in about 8.6K lines of Python and Triton~\cite{tillet2019triton} code. However, the scheduler and the cache manager are implemented in about 1.8K lines of C++ code, in order to accelerate request scheduling and scaling operations. For control message exchange between the master and worker models, we use the ZeroMQ~\cite{zeromq} messaging component. Transmission of KV cache across GPUs is managed using the \texttt{torch.distributed} library in PyTorch~\cite{paszke2019pytorch}.

In addition, SwiftCache integrates several common LLM inference optimization techniques, including continuous batching~\cite{yu2022orca}, FlashAttention~\cite{dao2022flashattention}, and PagedAttention~\cite{kwon2023efficient}. 
The current implementation supports the Qwen3~\cite{yang2025qwen3} series and LLaMA2~\cite{touvron2023llama} series models. %The code of SwiftCache is open-source and is publicly available at \url{https://anonymous.4open.science/r/SwiftCache-97FE}.
\begin{table}[t]
\centering
\caption{Model specs and per-token KV cache size (float16)}
\label{table:model_details}
\begin{tabular}{lcccc}
\toprule
{\small Model} & {\small Layers} & {\small Heads} & {\small Head Dim} & {\small Overhead/Token} \\
\midrule
LWM            & 32 & 32 & 128 & 0.50 MB \\
Qwen3-8B       & 36 & 8  & 128 & 0.141 MB \\
Qwen3-14B      & 40 & 8  & 128 & 0.156 MB \\
Qwen3-32B      & 64 & 8  & 128 & 0.25 MB \\
\bottomrule
\end{tabular}
\end{table}

\begin{figure*}[t]
  \centering
  \includegraphics[width=\textwidth]{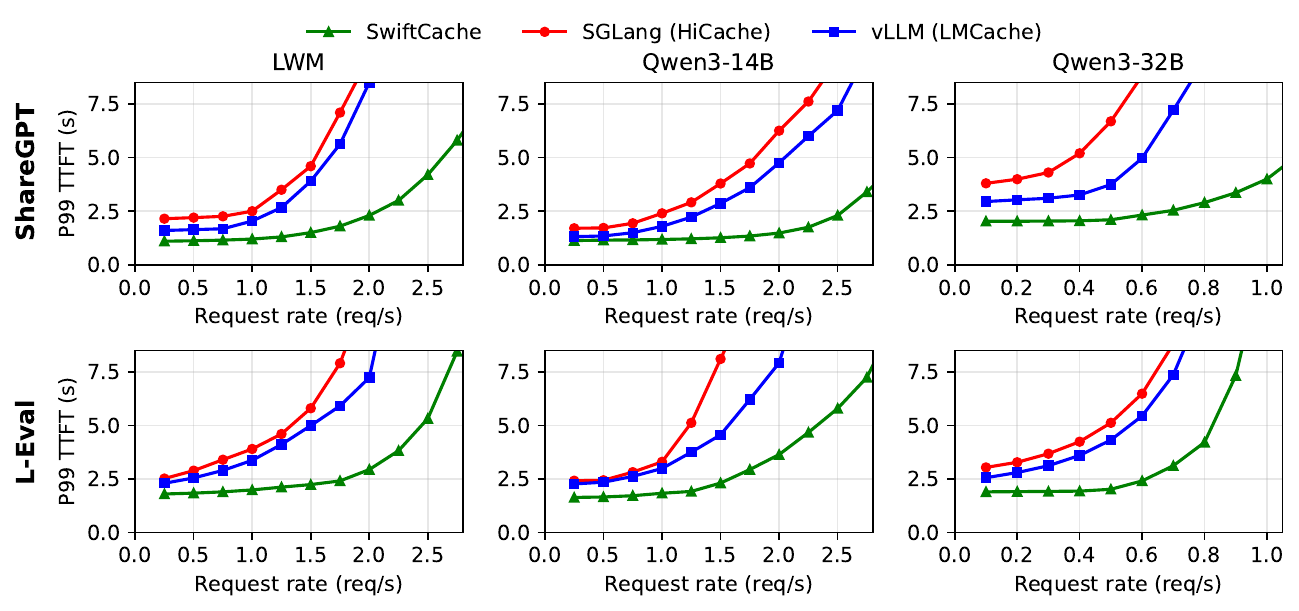}
\caption{P99 TTFT of SwiftCache vs. vLLM (LMCache) and SGLang (HiCache) on ShareGPT and L-Eval datasets.}
  \label{fig:p99_ttft}
  \vspace{-0.5em}
\end{figure*}
\section{Evaluations}\label{sec:evaluation}
In this section, we present our experimental setup and evaluate the performance of SwiftCache with state-of-the-art LLM inference systems under various real-world workloads.

\textbf{Model.} We use the Qwen3~\cite{yang2025qwen3} series models in our 
evaluation, including Qwen3-8B, Qwen3-14B, and Qwen3-32B. 
All three models support a maximum context length of 40K tokens and adopt the GQA~\cite{ainslie2023gqa} mechanism, which optimizes the size of the KV cache.  In addition, to highlight the impact of KV cache transmission on inference performance, we also use the LWM-1M-Text~\cite{liu2025world} model. This model shares the same parameter scale and architecture as Llama2-7B~\cite{touvron2023llama,touvron2023llama2}, employs the conventional MHA~\cite{vaswani2017attention} technique, and supports a maximum context length of 1 million tokens. All models use FP16 precision. The detailed specifications of these models are summarized in Table~\ref{table:model_details}.

\textbf{Testbed.} Our experiments are conducted on a server equipped with four NVIDIA H20 GPUs (96 GB each), a 64-core CPU, and 128 GB of host memory. The GPUs are interconnected via NVLink with a bidirectional bandwidth of 400 GB/s. The software environment includes Python 3.10.18, PyTorch 2.7, CUDA 12.4, and OpenAI Triton 3.3.0. The connection between the host memory and the GPUs is through PCIe 4.0, providing a total bandwidth of 32 GB/s that is shared among the four GPUs.

\textbf{Workloads.} In our evaluation, requests are sent to the server according to a Poisson arrival process, and all requests originate from the following two datasets.

\begin{itemize}
    \item \textbf{ShareGPT~\cite{sharegpt90k}}: It is a real-world dataset collected from user interactions with ChatGPT~\cite{openaiChatGPT}, and has been widely used in recent studies. It contains approximately 90K multi-turn conversations, with session length ranging from 12 to 602K tokens.

    \item \textbf{L-Eval~\cite{an2024eval}}: It is a comprehensive dataset with more than 500 tasks, covering long-context multi-turn conversations, question answering, and summarization tasks. The sequence length in this dataset ranges from 2.6K to 195.0K.
\end{itemize}

\textbf{Baselines.} We compare SwiftCache to three state-of-the-art baseline systems:
\begin{itemize}
    \item \textbf{vLLM~\cite{kwon2023efficient} (LMCache~\cite{lmcache,liu2025lmcache})\footnote{vLLM version is 0.14.0 and LMCache version is 0.3.15.}}: vLLM is an open-source serving engine for LLMs. LMCache stores KV pairs of computed prefixes to reduce redundant computation across requests. We use LMCache as implemented in vLLM as a representative cache-based serving baseline.
    
    \item \textbf{vLLM w/ chunked prefill~\cite{agrawal2024taming}}: vLLM with chunked prefill reduces peak GPU memory usage by processing prompts in chunks, thereby lowering activation memory overhead during prefill. We use it as a baseline for measuring maximum context length.
    
    \item \textbf{SGLang~\cite{zheng2024sglang} (HiCache)\footnote{SGLang version is 0.5.3, and HiCache is a built-in module in SGLang and does not have its own version number.}}: SGLang is an LLM serving framework designed for efficient multi-turn conversations and complex task execution. HiCache extends caching by maintaining hierarchical cache~\cite{gao2024cost,yu2025stateful} structures to maximize prefix reuse under diverse workloads. We adopt HiCache in SGLang as another representative caching baseline.
\end{itemize}

\textbf{Metrics.} We use the following metrics to evaluate the performance and effectiveness of SwiftCache: 
\begin{itemize}
    \item \textbf{P99 TTFT}: The 99\textsuperscript{th} percentile of the time-to-first-token latency, which reflects the worst-case responsiveness for the first output token across requests.
    \item \textbf{Interference}: It measures the interference experienced by the worker model, defined as the ratio of the inference latency without interference to the inference latency under interference.
    \item \textbf{Maximum Context Length}: The largest input sequence length that the system can handle, indicating the capability to process long-context requests.
\end{itemize}

\subsection{End-to-End Performance}
In this section, we conduct experiments using LWM, Qwen3-14B, and Qwen3-32B on the ShareGPT and L-Eval datasets. For each experiment, we designate one model as the master, with the remaining two serving as workers. We measure the master’s P99 TTFT, the performance degradation of each worker during the prefill and decoding stages, and the maximum request length that the master can handle. Both baselines are configured with 96\,GB of CPU memory as the offloading space.

Before conducting the formal experiments, we first send the initial rounds of multi-turn conversations to the server as a warm-up process, in order to eliminate the substantial recomputation overhead caused by the first requests and their impact on the P99 latency. 
To better demonstrate SwiftCache's performance in long-context conversations, for LWM, we select requests with lengths greater than 20K, with no upper limit, from each dataset. For Qwen3-14B and Qwen3-32B, we select requests ranging from approximately 20K to 40K tokens, as the maximum context length supported by the Qwen3 series is 40K.
As shown in Figure~\ref{fig:p99_ttft}, SwiftCache reduces P99 TTFT by approximately 67\% and 69\% on the ShareGPT and L-Eval datasets, respectively, compared to SGLang (HiCache). Compared to vLLM (LMCache), SwiftCache reduces P99 TTFT by approximately 54\% and 58\% on these two datasets, respectively. These improvements primarily stem from two optimizations: SwiftCache leverages high-speed NVLink instead of PCIe for KV cache transfers, and utilizes idle GPU memory on worker GPUs to expand the capacity of the prefix cache.
\begin{figure*}[t]
  \centering
  \includegraphics[width=0.95\textwidth]{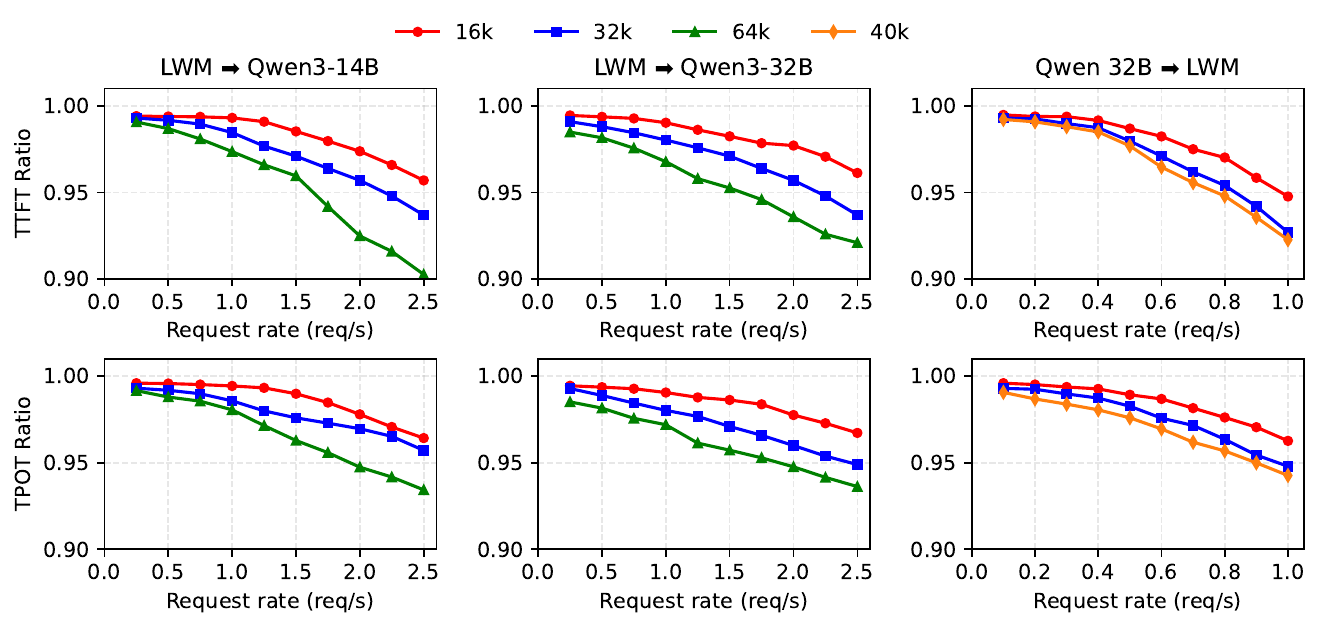}
\caption{Normalized TTFT and TPOT of different worker models under master interference in SwiftCache.}
  \label{fig:interference}
  \vspace{-0.5em}
\end{figure*}

\subsection{Interference on Worker Models}
We evaluate the interference from the master model on the worker models during the prefill and decoding stages under varying input lengths and request rates, as shown in Figure~\ref{fig:interference}. 
When LWM is used as the master, we test input lengths of 16K, 32K, and 64K tokens. For Qwen3-14B as the worker, the maximum increases in TTFT across these lengths reach 9.7\%, while Time-Per-Output-Token (TPOT) increases by up to 6.5\%. For Qwen3-32B as the worker, TTFT increases by up to 7.9\% and TPOT by up to 6.4\% across the same three lengths. 
Given that Qwen3-32B has a built-in maximum inference length of 40K tokens, we evaluate it with input lengths of 16K, 32K, and 40K tokens when serving as the master. In this setting, TTFT increases by up to 7.7\% and TPOT by up to 5.6\%. 

These results demonstrate that in SwiftCache, interference from the master to the worker is minimal. This is because KV cache loading and storing operations on worker GPUs do not occupy any compute resources, consuming only network resources and a negligible fraction of GPU memory bandwidth. The network resource usage between the master and worker has no impact on worker performance. Overall, SwiftCache delivers substantial performance gains for the master, but it incurs only a slight performance loss on the workers ($\leq$9.7\%).

\subsection{Maximum Context Length}
In model inference, the achievable maximum context length is often constrained by the memory capacity of the GPU on which the model is running. In our experiments, we selected two categories of GPUs to represent typical industrial inference environments: high capacity GPUs with 96GB and 80GB VRAM, corresponding to NVIDIA H20 and A100, and low capacity GPUs with 32GB and 24GB VRAM, corresponding to NVIDIA V100 and L4. For the 96GB, 80GB, and 32GB settings, we chose Qwen3-14B and Qwen3-8B as worker models; for the 24GB setting, we deployed two Qwen3-8B workers, as Qwen3-14B cannot fit into 24GB memory. To more comprehensively evaluate the effectiveness of SwiftCache, we do not consider the maximum context length specified in the model’s configuration.

\begin{figure*}[t]
  \centering
  \includegraphics[width=\textwidth]{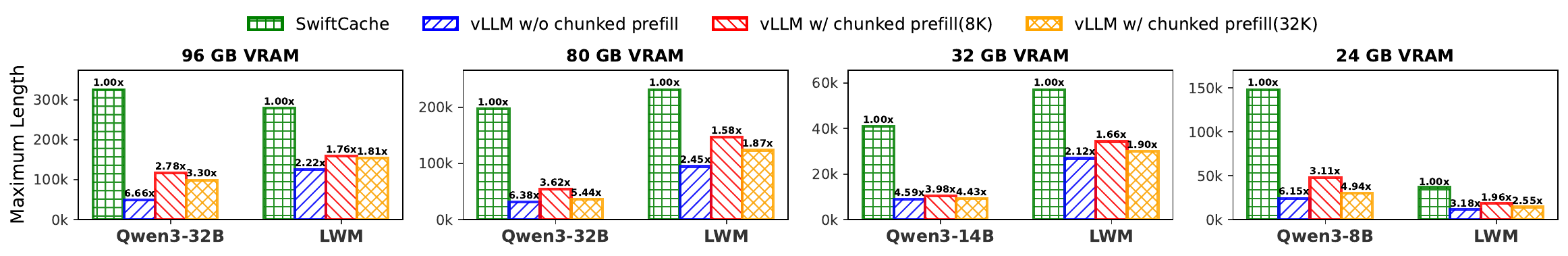}
\caption{Maximum inference length supported by SwiftCache and vLLM (without chunked prefill, and with chunked prefill using chunk sizes of 8K and 32K) across different VRAM capacities and model sizes.} 
  \label{fig:max_length}
  \vspace{-0.5em}
\end{figure*}

Figure~\ref{fig:max_length} shows the maximum context length achievable by the master model under these configurations. Among the three baselines, vLLM with chunked prefill~\cite{agrawal2024taming} using a chunk size of 8K tokens (the default configuration of vLLM) supports the longest context length, due to its smallest activation footprint during the prefill stage. However, SwiftCache achieves significantly longer contexts compared to this baseline, with improvements ranging from 1.58$\times$ to 3.98$\times$ across the tested VRAM settings. Compared to the other two baselines, vLLM without chunked prefill and vLLM with chunked prefill (chunk size = 32K), SwiftCache delivers memory efficiency gains corresponding to maximum context length improvements in the range of 1.81$\times$ to 6.66$\times$.

These results demonstrate that SwiftCache’s superiority in context length scalability stems from its \textit{layer stream cache} mechanism, described in \S~\ref{sec:master_cache_manager}, which enables efficient streaming of layer outputs and reduces peak activation memory usage during inference. Moreover, LSC is orthogonal to chunked prefill, seamlessly integrating with it to further alleviate activation memory pressure in ultra-long contexts.

\begin{figure}[t]
  \centering
  \includegraphics[width=\columnwidth]{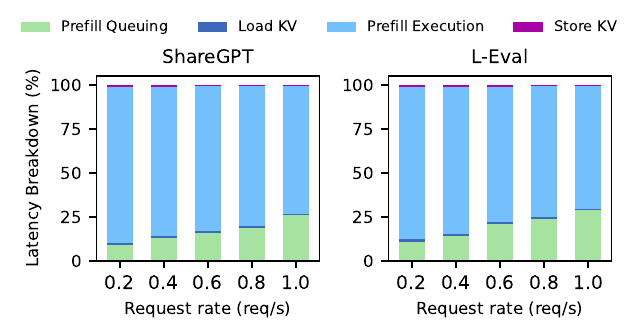}
\caption{Left: The latency breakdown when serving Qwen3-32B on the ShareGPT dataset with SwiftCache. Right: The latency breakdown when serving Qwen3-32B on the L-Eval dataset with SwiftCache.}
  \label{fig:breakdown}
  \vspace{-1em}
 \end{figure}

 \subsection{Latency Breakdown}
To understand SwiftCache’s effectiveness in more detail, we perform a latency breakdown of the prefill stage. We divide the lifecycle of the prefill stage into four phases: \textit{prefill queuing}, \textit{load KV}, \textit{prefill execution}, and \textit{store KV}. The total latency of the prefill stage is obtained by summing the time spent in each phase across all requests. In this analysis, we temporarily ignore any overlap between computation and KV cache transmission.

Figure~\ref{fig:breakdown} shows the latency breakdown for the Qwen3-32B model on the ShareGPT and L-Eval datasets. We chose Qwen3-32B because KV cache transmission is more demanding for larger models. Despite this, KV cache loading and storing account for less than 1.4\% and 1.1\% of the total latency, respectively. When considering the overlap between computation and transmission, the time spent on KV cache transmission drops to below 0.1\%. 

\begin{figure}[t]
  \centering
  \includegraphics[width=\columnwidth]{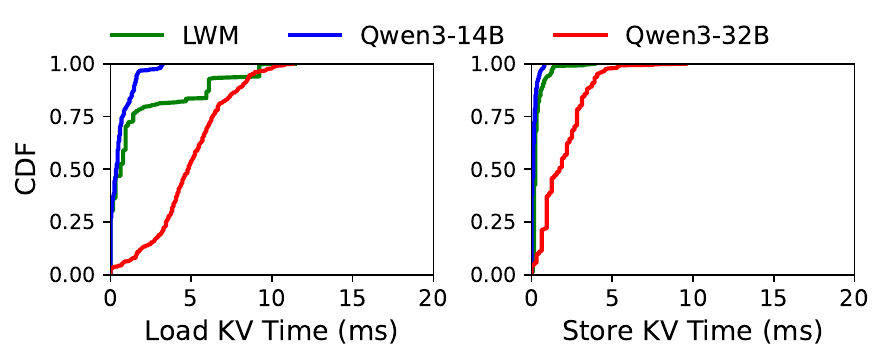}
\caption{Left: The CDF of KV loading time for three models. Right: The CDF of KV storing time for three models.}
  \label{fig:cdf}
   \vspace{-1em}
 \end{figure}

By examining the CDFs of transmission delay in Figure~\ref{fig:cdf}, we observe that even for Qwen3-32B, 99\% of requests have KV cache loading times under 11\,ms, and 99\% of requests have KV cache storing times under 6\,ms. Such overhead is negligible for Qwen3-32B requests with more than 10K tokens, where computation time is orders of magnitude greater, ranging from hundreds of milliseconds to several seconds. Although LWM has a larger KV cache footprint per token as shown in Table~\ref{table:model_details}, its significantly smaller model weights compared to Qwen3-32B leave more GPU memory available for \textit{regular cache} allocation. Consequently, more KV pairs can be stored in the local \textit{regular cache} rather than workers' \textit{elastic cache}, thereby reducing transmission overhead.

\section{Related Work}
\textbf{Efficient LLM serving}: Recently, many studies~\cite{lin2024parrot,hu2025deepserve,wu2023fast,qiu2025modserve,zhu2025nanoflow} have focused on improving the inference efficiency of LLM serving systems. Orca~\cite{yu2022orca} introduces \textit{continuous batching}, which eliminates the need for padding between requests of different lengths, and leverages iteration-level scheduling to remove completed requests from the inference engine promptly. vLLM~\cite{kwon2023efficient} proposes \textit{PagedAttention} to reduce memory fragmentation and improve memory utilization, thereby increasing the throughput of LLM serving. SwiftCache integrates both continuous batching and PagedAttention to enhance general inference performance for LLMs. DistServe~\cite{zhong2024distserve} and Splitwise~\cite{patel2024splitwise} decouple the \textit{prefill} and \textit{decoding} stages to meet the SLO requirements of each stage individually. SwiftCache is also compatible with such PD separation strategies.

\noindent \textbf{KV cache reusing}: CachedAttention~\cite{gao2024cost,yu2025stateful, hu2026bidaw} extends the capacity of KV cache to host memory and SSD storage, enabling the system to support more prefixes and reduce recomputation overhead during the prefill stage. Mooncake~\cite{qin2025mooncake} constructs a centralized KV cache pool shared across multiple model instances, thereby improving throughput in multi-turn dialogue scenarios. SwiftCache dynamically extends the KV cache of a model to the GPUs of other models, which, compared to existing PCIe-based KV cache transmission, substantially reduces transfer latency and improves overall GPU memory utilization.

In addition, many works~\cite{lee2024infinigen,yao2025cacheblend,liu2026cacheslide,zheng2026solidattention,sheng2023flexgen} exploit the sparsity in the KV cache or decoding layers to enable cache reuse. While these approaches can significantly increase system throughput, they may also cause semantic accuracy degradation due to partial cache utilization.

Notably, both Mooncake and DroidSpeak~\cite{liu2024droidspeak} involve sharing KV cache across LLMs. Mooncake shares KV cache across different instances of the same model, whereas DroidSpeak supports LLMs originating from the same foundation models, since such models must have identical architectures. In contrast, SwiftCache enables direct sharing of KV cache memory buffers across heterogeneous models, even when their architectures differ.

\noindent \textbf{Long context inference:}  
PrefillOnly~\cite{du2025prefillonly} proposes \textit{hybrid prefilling} to reduce activations during inference, thereby lowering memory usage and supporting longer text generation. LoongServe~\cite{wu2024loongserve} introduces \textit{elastic sequence parallelism}, which reduces KV cache transmission overhead and efficiently supports long-context inference. Chunked prefill further improves inference efficiency by splitting long prompts into smaller chunks during prefill, reducing peak memory usage and enabling support for longer sequences. In addition, works like~\cite{adnan2024keyformer,tang2024quest,xiao2024efficient} reduce the size of the KV cache to lower memory consumption and further support longer text generation, but at the cost of some accuracy degradation, whereas SwiftCache does not incur such loss.

\section{Conclusions and Discussions}
In this paper, we presented SwiftCache, a multi-model collaborative LLM serving system for efficient multi-turn conversations with heterogeneous KV cache sharing. By leveraging NVLink-based cache sharing and layer stream cache together with elastic cache, SwiftCache improves memory utilization and serving efficiency for long-context inference. Experimental results also show that SwiftCache reduces P99 TTFT by up to 69\% and extends the maximum context length by up to 3.98× compared with state-of-the-art baselines, while incurring only modest interference to co-located models. These results demonstrate the effectiveness of SwiftCache for efficient long-context multi-turn LLM serving. Currently, SwiftCache is primarily designed for deployment on a single multi-GPU server, since GPUs located on different servers cannot communicate directly through NVLink. Extending SwiftCache to multi-server environments would incur significant communication overhead and may require additional hardware support, which we leave for future work.

\section*{Code Availability}
The code of SwiftCache is open-source and publicly available at https://anonymous.4open.science/r/SwiftCache-97FE for research usage.

% \begin{acks}
% \end{acks}

\bibliographystyle{ACM-Reference-Format}
\bibliography{references}

\end{document}